\documentclass[%
 reprint,
 amsmath,amssymb,
 aps,
]{revtex4-2}

\usepackage{graphicx}
\usepackage{dcolumn}
\usepackage{bm}

\usepackage{color}

\renewcommand{\v}[1]	{\ensuremath{\mathbf{#1}}} 
\newcommand{\mr}[1]     {\ensuremath{\mathrm{#1}}}
\newcommand{\mol}[1]	{\ensuremath{_{\text{#1}}}}

\begin{document}

\preprint{N/A}

\title{Examining the order-of-limits problem and lattice constant performance of the Tao--Mo Functional}

\author{James W. Furness$^1$*, Niladri Sengupta$^2$, Jinliang Ning$^1$, Adrienn Ruzsinszky$^2$, Jianwei Sun$^1$*}
 \affiliation{$^1$Department of Physics and Engineering Physics, Tulane University, New Orleans, LA 70118 \\
    $^2$Department of Physics, Temple University, Philadelphia, PA 19122
 }%
\email{jsun@tulane.edu}
\email{jfurness@tulane.edu}

\date{\today}

\begin{abstract}
In their recent communication [Phys. Rev. Lett., 117, 073001 (2016)]  Tao and Mo presented a semi-local density functional derived from the density matrix expansion of the exchange hole localised by a general coordinate transformation. We show that the order-of-limits problem present in the functional, dismissed as harmless in the original publication, causes severe errors in predicted phase transition pressures. We also show that the claim that lattice volume prediction accuracy exceeds that of existing similar functionals was based on comparison to reference data that misses anharmonic zero-point expansion and consequently overestimates accuracy. By highlighting these omissions, we give a more accurate assessment of the Tao-Mo functional and show a simple route to resolving the problems.
\end{abstract}

\maketitle


\section{\label{sec:intro} Introduction}

Many of the advances that have developed the accuracy of Kohn--Sham density functional theory (KS-DFT) have been realised by designing approximations to the exchange-correlation (XC) energy functional from theoretical analysis. This non-empirical approach to functional design is commonly pursued by obeying conditions known for the theoretical exact XC functional, termed exact constraints, and the resulting functionals have enjoyed broad success\cite{Kohn1965, Vosko1980, Perdew1996, Cancio2018, Tao2003, Sun2015}. A complementary approach to functional design, though less well explored in recent years, has been to derive functionals from models of the exact exchange hole \cite{Becke1989, Tao2016}. A recent advance in this approach was made by Tao and Mo in Ref. \cite{Tao2016}, in which a new semi-local density functional approximation for the exchange energy was derived from a general coordinate transformation \cite{Perdew2007} to the density matrix expansion of the exact exchange hole.

The Tao--Mo exchange hole model was used to construct a meta-generalised gradient approximation (meta-GGA) exchange functional from the electron density, electron density gradient, and the orbital kinetic energy density, $\tau(\v r) = 1/2 \sum_i^{\mr{occ.}}|\nabla\psi_i(\v r)|^2$. The resulting exchange energy density is combined with a modified TPSS correlation functional \cite{Tao2003} with simplified spin polarisation and re-parametrised to better fit the exact exchange correlation energy of the one electron Gaussian density. The resulting functional was denoted ``TM''. Combination of the new exchange functional with unmodified TPSS correlation was also suggested and named ``TMTPSS''.

The resulting non-empirical meta-GGA TM functional properly recovers the uniform electron gas, the slowly-varying density limit, and the iso-orbital limits. Its useful accuracy was established in Ref. \cite{Tao2016} against equilibrium 0 Kelvin lattice constants of 16 solids, alongside atomisation energies, Jellium surface energies, dissociation energies of hydrogen bonded complexes, and cohesive energies of solids.

While the TM functional presents an intriguing advance for building functionals from exchange hole models, it contains fundamental issues that limit its accuracy for some classes of problems. Here we examine the order-of-limits problem and its impact on phase transition pressure prediction. We also show that the TM functional's accuracy for lattice constants was overestimated in Ref. \cite{Tao2016} due to the missing zero-point expansion (ZPE) correction in the reference data used.

\section{\label{sec:order} Order-of-limits Problem}

Like the earlier TPSS exchange functional\cite{Tao2003}, TM uses the dimensionless meta-GGA indicator variable,
\begin{equation}
z(\v r) = \frac{\tau^{\mr{vW}}(\v r)}{\tau(\v r)},
\end{equation}
where $\tau^{\mr{vW}}(\v r) = |\nabla n(\v r)|^2/(8n(\v r))$ is the single orbital limit for the kinetic energy. In TM, $z(\v r)$ is used to identify single orbital and slowly varying densities through the interpolation function, 
\begin{equation}
    w(\v r) = \frac{z(\v r)^2 3z(\v r)^3}{(1 + z(\v r)^3)^2}.
    \label{eq:ie}
\end{equation}
A different iso-orbital indicator,
\begin{equation}
    \alpha(\v r) = \frac{\tau(\v r) - \tau^{\mr{vW}}(\v r)}{\tau^{\mr{UEG}}(\v r)},
\end{equation}
where $\tau^{\mr{UEG}}(\v r) = (3/10)(3\pi^2)^{2/3}n(\v r)^{5/3}$ is the kinetic energy density of the uniform electron gas, occurs in the terms that recover the fourth-order gradient expansion of the exchange energy. These indicators are related as,
\begin{align}
    z(\v r) &= \frac{1}{1+\frac{3\alpha(\v r)}{5p(\v r)}},\label{eq:z} \\ 
    \alpha(\v r) &= \frac{5p(\v r)}{3}\left[ \frac{1}{z(\v r)} - 1 \right], \label{eq:a_z}
\end{align}
through square of the reduced density gradient,
\begin{align}
    p(\v r) = s(\v r)^2 &= \frac{|\nabla n(\v r)|^2}{4(3\pi^2)^{2/3}n(\v r)^{8/3}}, \label{eq:p} \\
    &= \frac{3}{5}\frac{\tau^{\mr{vW}}(\v r)}{\tau^{\mr{UEG}}(\v r)}. \label{eq:p_t}
\end{align}
Other iso-orbital indicator functions are also known\cite{Zhao2006, Furness2019}.

Combined dependence on $\alpha$ and $z$ introduces an order-of-limits problem into TM that was first identified for TPSS in Ref. \cite{Perdew2004}. This problem was identified in the initial publication of TM \cite{Tao2016} but disregarded as harmless. To the contrary, the order-of-limits problem has been shown to be the leading cause of error in TPSS predictions of phase transition pressures\cite{Ruzsinszky2012} and we will show here that the same is true for TM.

\begin{figure}
    \includegraphics[width=\linewidth]{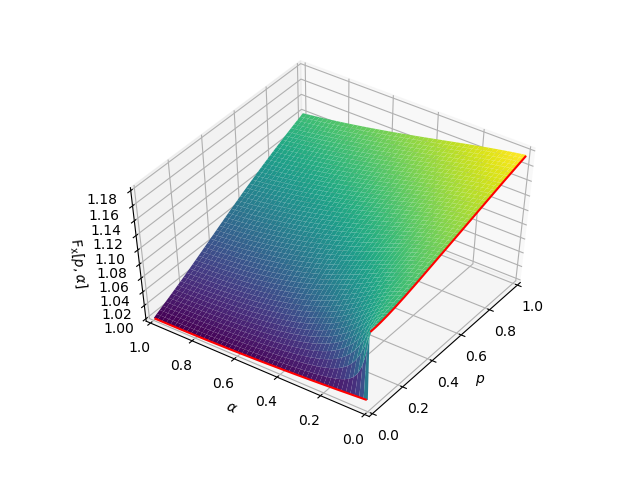}
    \caption{The exchange enhancement factor for the TM functional of Ref. \cite{Tao2016} as a function of $p$ and $\alpha$. The order-of-limits discontinuity is visible at $(0, 0)$. Edges $p = 0$ and $\alpha = 0$ are highlighted red.}
    \label{fig:surf}
\end{figure}

The order-of-limits discontinuity can be seen when the enhancement factor, $F_{\mr{x}}$, (Eq. 11 of Ref. \cite{Tao2016}) is expressed in terms of $p$ and $\alpha$ using Eqs. \ref{eq:z}-\ref{eq:p_t}. Taking the limit of $p \rightarrow 0$ followed by the limit $\alpha \rightarrow 0$ gives,
\begin{equation}
    \lim_{\alpha \rightarrow 0}\left[ \lim_{p\rightarrow0} \left[F_\mr{x}(p, \alpha)\right]\right] = 1.0137,
\end{equation}
whereas reversing the order and taking the limit of $\alpha \rightarrow 0$ followed by $p \rightarrow 0$,
\begin{equation}
    \lim_{p \rightarrow 0}\left[ \lim_{\alpha\rightarrow0} \left[F_\mr{x}(p,\alpha)\right]\right] = 1.1132.
\end{equation}
This discontinuity is shown graphically in Figure \ref{fig:surf}, which plots the TM exchange enhancement as a function of $p$ and $\alpha$. Following the $F_{\mr{x}}[p=0,\alpha]$ and $F_{\mr{x}}[p, \alpha=0]$ edges, highlighted red, reveals the discontinuity at $F_{\mr{x}}[p=0, \alpha=0]$.

\begin{figure}
    \centering
    \includegraphics[width=\linewidth]{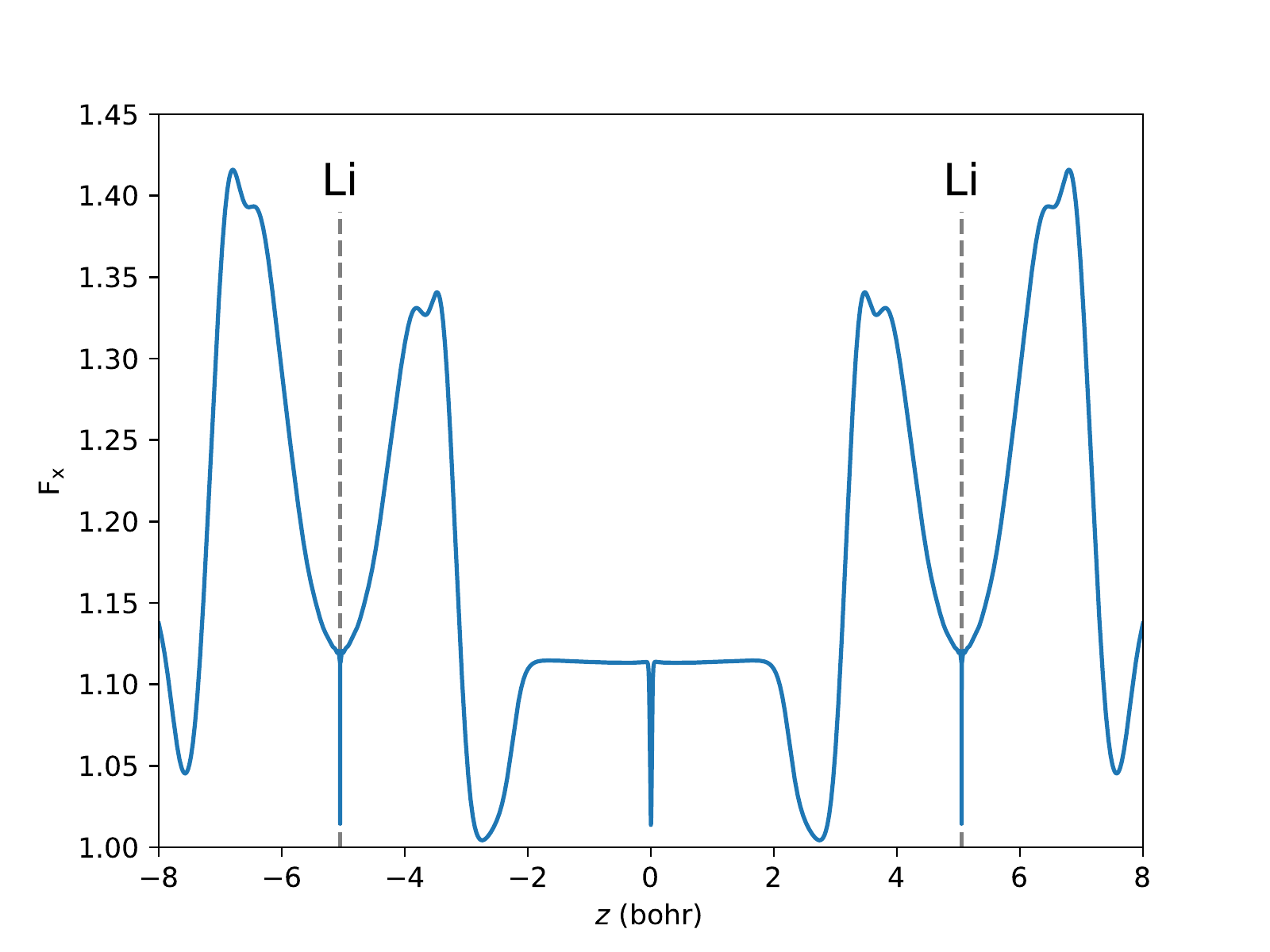}
    \caption{The exchange enhancement factor of the TM functional\cite{Tao2016} along the Li\mol{2} internuclear axis, evaluated for Hartree--Fock orbitals in the aug-cc-pVQZ basis\cite{Kendall1992}. Order-of-limits discontinuities are visible at the nuclei (indicated with dashed vertical lines) and at the bond centre $z = 0$.}
    \label{fig:li2}
\end{figure}

Ref. \cite{Tao2016} asserts that the discontinuity at $p = \alpha = 0$ is not a practical concern, stating that such behaviour only occurs close to the nuclei. This assertion is incorrect and important counter examples are found at the centre of stretched covalent single bonds\cite{Perdew2004, Ruzsinszky2012, Xiao2013}. An example of this is shown in Figure \ref{fig:li2}, which plots $F_{\mr{x}}$ along the bond axis of stretched Li\mol{2}. The effect of the order-of-limits discontinuity is clearly seen at the bond centre (as well as at the nuclei) as downwards spikes caused by the exchange enhancement jumping between the two limits.

Far from being harmless, the order-of-limits problem was identified in Refs. \cite{Ruzsinszky2012} and \cite{Xiao2013} as the leading source of errors for TPSS in crystal structure energy differences and in the cohesive energies of insulating solids. The severity of this problem is clearly revealed by comparing the accuracy of a small but representative set of phase transitions calculated by TM, that suffers the order-of-limits problem) with those calculated of the SCAN functional \cite{Sun2015} which does not. 

The test set is comprised of semiconductor-metal, metal-metal, and semiconductor-insulator transitions for which thermal effects are small enough to allow direct comparison between experimental and predicted results. All calculations were made following the procedure of Ref. \cite{Sengupta2018}, using the Vienna ab initio simulation package (VASP) \cite{Kresse1993, Kresse1996} with PBE projector augmented waves that include the kinetic energy density component for B, N, Si, O, and C and fully occupied $d$-shell electrons with kinetic energy density components for Ga, As, Pb, and Ge. Gamma-centered Monkhorst--Pack k-mesh were used throughout. The k-mesh point densities and cutoff energy for the plane wave basis are as detailed in Ref. \cite{Sengupta2018} supplemental material Table S1.

First, we note that both TM and SCAN give high accuracy equilibrium cell volumes for all phases of the test systems. The similar performance for equilibrium volumes follows directly from accurate lattice constant predictions which are not directly affected by the order-of-limits problem. A different result is seen for the transition pressures, which are reported in Table \ref{tab:tp}. The order-of-limits problem causes the TM functional to strongly underestimate the phase transition energy for every system in the set, with qualitatively incorrect phase ordering seen for BN. A similar performance for TPSS was reported in Ref \cite{Xiao2013}, which has the same order-of-limits problem. In contrast, the SCAN functional which does not suffer the order-of-limits problem predicts accurate phase transition pressures for every system. We must therefore conclude that, contrary to Ref. \cite{Tao2016}, the order-of-limits problem in TM cannot be dismissed.

\begin{table}
    \caption{Equilibrium cell volumes ($\text{\AA}{}^3$). Mean error (ME) and mean absolute error (MAE) are given relative to experiment. Experimental reference values were taken from the reported ICSD values via the Materials Project database\cite{Jain2013} unless otherwise noted. Computational details are given in main text.}
    \label{tab:eq_vol}
    \centering
    \begin{tabular}{lrrr}
        \hline
        \hline
         & SCAN & TM & Expt.\cite{Jain2013} \\
        \hline
        Si(Diamond) & 39.976 & 39.623 & 40.037 \\
        Si($\beta$Sn) & 59.772 & 58.583 & 55.82$^*$ \\
        Ge(Diamond) & 45.287 & 43.902 & 45.271 \\
        Ge($\beta$Sn) & 74.214 & 68.925 & - \\ 
        GaAs(ZnS) & 45.592 & 45.414 & 45.138 \\
        GaAs(Cmcm) & 147.248 & 147.922 & - \\
        SiO2(Quartz) & 37.357 & 36.819 & 37.803 \\
        SiO2(Stishovite) & 23.284 & 23.590 & 23.325 \\
        Pb(FCC) & 30.648 & 29.949 & 30.010 \\
        Pb(HCP) & 60.976 & 59.591 & 48.530 \\
        BN(Cubic) & 11.742 & 11.763 & 11.664 \\
        BN(Hexagonal) & 35.963 & 35.108 & 36.701 \\
        \hline
        ME & 1.630 & 1.004 \\
        MAE & 1.887 & 1.889 \\
        \hline
        \hline
         & & & $^*$Ref. \cite{McMahon1994}
    \end{tabular}
\end{table}

\begin{table}
    \caption{Transition pressures (GPa) for structural phase transitions. Mean error (ME) and mean absolute error (MAE) are given relative to experiment\cite{Mujica2003}. Computational details given in main text.}
    \label{tab:tp}
    \centering
    \begin{tabular}{lrrr}
        \hline
        \hline
         & SCAN & TM & Expt.\cite{Mujica2003} \\
        \hline
        Si & 14.5 & 3.9 & 12.0 \\
        Ge & 11.3 & 6.7 & 10.6 \\
        GaAs & 17.1 & 8.2 & 15.0 \\
        SiO\mol{2} & 4.6 & 1.0 & 7.5$^*$\\
        Pb & 16.4 & 10.0 & 14.0 \\
        BN & 2.8 & -1.2 & 5.0 \\
        \hline
        ME & 0.4 & -5.9 \\
        MAE & 2.1 & 5.9 \\
        \hline
        \hline
         & & & $^*$Ref. \cite{Hamann1996}
    \end{tabular}
\end{table}

In principle, the order-of-limits problem could be removed from the TM functional to make a revised-TM functional analogous to the regularised-TPSS proposed in Ref. \cite{Ruzsinszky2012}. The order-of-limits problem stems from the interpolation function, Eq. \ref{eq:ie}, that joins the slowly varying exchange enhancement factor to the density matrix expansion exchange enhancement factor\cite{Tao2016}. If this interpolation function were substituted for a function of $\alpha$ under the constraint that,
\begin{align}
    w(\alpha = 1) &= 0
\end{align}
then the order-of-limits problem would be resolved. The fourth order gradient correction term, $F_\mr{x}^{\mr{SC}}$ in Ref. \cite{Tao2016}, should be suitably modified to maintain the correct gradient expansion for the new interpolation function.

As there is no simple mapping between $z$ and $\alpha$ and functional performance is likely to be sensitive to the exact nature of $w(\alpha)$, deriving and testing a revised TM functional is beyond the scope of the current communication. Inspiration for possible $w(\alpha)$ could be taken from other non-empirical interpolation based meta-GGA functionals\cite{Sun2013, Sun2013a, Sun2015, Sun2015a}. 

\section{\label{sec:lattice} Anharmonic Correction to Lattice Constants}

Accurate prediction of lattice constants is an important indicator of functional performance both as a measure directly relevant to experiment, and as a property that underpins many others. Direct comparison of calculated lattice constants to experimental data is complicated by zero-point phonon effects in the experimental data that cause an anharmonic zero-point expansion (ZPE) of measured lattice constants. The impact of this anharmonic ZPE was calculated in Ref. \cite{Hao2012} and found to expand lattice constants by around $0.015\text{\AA}$ $(\approx 0.35\%)$ for a set of 24 solids. 

The experimental reference data used in Ref. \cite{Tao2016} were obtained by extrapolating finite temperature experimental lattice constants to 0 Kelvin. Extrapolating in this way implicitly includes anharmonic ZPE effects so such data is not directly comparable with single point electronic structure calculations in which the nuclei are treated with harmonic potentials. Hence, neglecting to control for anharmonic ZPE introduced a systematic error into the assessment of TM performance for lattice constants.

The original assessment in Ref. \cite{Tao2016} was made from 13 bulk crystalline solids. This set includes main-group metals (Li, Al), semiconductors (diamond, Si, $\beta$-SiC, GaAs), ionic crystals (NaCl, NaF, LiCl, LiF, MgO), and transition metals (Cu, Ag). Comparable SCAN data is available from Ref. \cite{Sengupta2018} and is included here. We repeat the analysis of Ref. \cite{Tao2016} using reference data properly corrected for anharmonic ZPE, the results of which are presented in Table \ref{tab:lat_con}. Computational details follow those detailed in Section \ref{sec:order}.

The uncorrected reference data used in Ref. \cite{Tao2016} indeed suggests the conclusion that TM, and to a lesser extent TMTPSS, predicts lattice constants with higher accuracy than SCAN. The reference data with proper anharmonic ZPE corrections shows the error of this conclusion however, with SCAN instead having higher accuracy than both TM and TMTPSS.

\begin{table}
    \caption{Mean error (ME) and mean absolute error (MAE) for lattice constants ($\text{\AA}$) of 13 bulk crystalline solids calculated with zero-point expansion (ZPE) correction, and without (as used in Ref. \cite{Tao2016}).}
    \label{tab:lat_con}
    \centering
    \begin{tabular}{l|rr|rr}
         & \multicolumn{2}{|c}{ZPE-Uncorrected} & \multicolumn{2}{|c}{ZPE-Corrected} \\
         & ME & MAE & ME & MAE \\
         \hline
        SCAN & -0.013 & 0.018 & 0.004 & 0.011 \\
        TM & -0.001 & 0.012 & 0.015 & 0.019 \\
        TMTPSS & 0.008 & 0.015 & 0.024 & 0.028
    \end{tabular}
\end{table}

\section{\label{sec:conclusion} Conclusion}

Whilst Ref. \cite{Tao2016} presents an appealing non-empirical functional, assessment of its performance was flawed in two important aspects. Firstly, the order-of-limits problem is more severe for TM than initially claimed and we have shown that it causes significant error for transition pressure predictions. A route to revising the functional to remove this problem is clear however, if the interpolation function is redesigned as a function of $\alpha$. Secondly, whilst both TM and TMTPSS make accurate predictions of lattice constants, the lack of anharmonic ZPE correction in the reference data caused Ref. \cite{Tao2016} to incorrectly conclude that TM is more accurate than other meta-GGA functionals, such as SCAN, for this property. When the assessment is repeated with properly ZPE-corrected reference data the apparent accuracy is worsened with TM and TMTPSS showing worse accuracy than SCAN, though we stress that all three functionals are impressively accurate. Given the compelling theoretical foundations of TM, we feel that the functional would be well served by a revision that solves the order-of-limits problem. 

\section{Acknowledgements}
J. F., J. N., and J. S. acknowledge the support of the U.S. DOE, Office of Science, Basic Energy Sciences Grant No. DE-SC0019350 (core research). The research of  A. R. was supported by the National Science Foundation under Grant No.DMR-1553022. The authors were saddened to hear of the recent passing of Prof. J. Tao and would like to recognise the valuable contributions he made to the DFT community and the wider sciences.

\section{Data Availability Statement}
The data that support the findings of this study are available from the corresponding author upon reasonable request.

\bibliography{TM_Comment}

\begin{thebibliography}{26}%
\makeatletter
\providecommand \@ifxundefined [1]{%
 \@ifx{#1\undefined}
}%
\providecommand \@ifnum [1]{%
 \ifnum #1\expandafter \@firstoftwo
 \else \expandafter \@secondoftwo
 \fi
}%
\providecommand \@ifx [1]{%
 \ifx #1\expandafter \@firstoftwo
 \else \expandafter \@secondoftwo
 \fi
}%
\providecommand \natexlab [1]{#1}%
\providecommand \enquote  [1]{``#1''}%
\providecommand \bibnamefont  [1]{#1}%
\providecommand \bibfnamefont [1]{#1}%
\providecommand \citenamefont [1]{#1}%
\providecommand \href@noop [0]{\@secondoftwo}%
\providecommand \href [0]{\begingroup \@sanitize@url \@href}%
\providecommand \@href[1]{\@@startlink{#1}\@@href}%
\providecommand \@@href[1]{\endgroup#1\@@endlink}%
\providecommand \@sanitize@url [0]{\catcode `\\12\catcode `\$12\catcode
  `\&12\catcode `\#12\catcode `\^12\catcode `\_12\catcode `\%12\relax}%
\providecommand \@@startlink[1]{}%
\providecommand \@@endlink[0]{}%
\providecommand \url  [0]{\begingroup\@sanitize@url \@url }%
\providecommand \@url [1]{\endgroup\@href {#1}{\urlprefix }}%
\providecommand \urlprefix  [0]{URL }%
\providecommand \Eprint [0]{\href }%
\providecommand \doibase [0]{https://doi.org/}%
\providecommand \selectlanguage [0]{\@gobble}%
\providecommand \bibinfo  [0]{\@secondoftwo}%
\providecommand \bibfield  [0]{\@secondoftwo}%
\providecommand \translation [1]{[#1]}%
\providecommand \BibitemOpen [0]{}%
\providecommand \bibitemStop [0]{}%
\providecommand \bibitemNoStop [0]{.\EOS\space}%
\providecommand \EOS [0]{\spacefactor3000\relax}%
\providecommand \BibitemShut  [1]{\csname bibitem#1\endcsname}%
\let\auto@bib@innerbib\@empty
\bibitem [{\citenamefont {Kohn}\ and\ \citenamefont {Sham}(1965)}]{Kohn1965}%
  \BibitemOpen
  \bibfield  {author} {\bibinfo {author} {\bibfnamefont {W.}~\bibnamefont
  {Kohn}}\ and\ \bibinfo {author} {\bibfnamefont {L.~J.}\ \bibnamefont
  {Sham}},\ }\bibfield  {title} {\bibinfo {title} {{Self-consistent equations
  including exchange and correlation effects}},\ }\href
  {https://doi.org/10.1103/PhysRev.140.A1133} {\bibfield  {journal} {\bibinfo
  {journal} {Physical Review}\ }\textbf {\bibinfo {volume} {140}},\ \bibinfo
  {pages} {A1133} (\bibinfo {year} {1965})}\BibitemShut {NoStop}%
\bibitem [{\citenamefont {Vosko}\ \emph {et~al.}(1980)\citenamefont {Vosko},
  \citenamefont {Wilk},\ and\ \citenamefont {Nusair}}]{Vosko1980}%
  \BibitemOpen
  \bibfield  {author} {\bibinfo {author} {\bibfnamefont {S.~H.}\ \bibnamefont
  {Vosko}}, \bibinfo {author} {\bibfnamefont {L.}~\bibnamefont {Wilk}},\ and\
  \bibinfo {author} {\bibfnamefont {M.}~\bibnamefont {Nusair}},\ }\bibfield
  {title} {\bibinfo {title} {{Accurate spin-dependent electron liquid
  correlation energies for local spin density calculations: a critical
  analysis}},\ }\href {https://doi.org/10.1139/p80-159} {\bibfield  {journal}
  {\bibinfo  {journal} {Canadian Journal of Physics}\ }\textbf {\bibinfo
  {volume} {58}},\ \bibinfo {pages} {1200} (\bibinfo {year}
  {1980})}\BibitemShut {NoStop}%
\bibitem [{\citenamefont {Perdew}\ \emph {et~al.}(1996)\citenamefont {Perdew},
  \citenamefont {Burke},\ and\ \citenamefont {Ernzerhof}}]{Perdew1996}%
  \BibitemOpen
  \bibfield  {author} {\bibinfo {author} {\bibfnamefont {J.~P.}\ \bibnamefont
  {Perdew}}, \bibinfo {author} {\bibfnamefont {K.}~\bibnamefont {Burke}},\ and\
  \bibinfo {author} {\bibfnamefont {M.}~\bibnamefont {Ernzerhof}},\ }\bibfield
  {title} {\bibinfo {title} {{Generalized Gradient Approximation Made
  Simple.}},\ }\href@noop {} {\bibfield  {journal} {\bibinfo  {journal}
  {Physical Review Letters}\ }\textbf {\bibinfo {volume} {77}},\ \bibinfo
  {pages} {3865} (\bibinfo {year} {1996})}\BibitemShut {NoStop}%
\bibitem [{\citenamefont {Cancio}\ \emph {et~al.}(2018)\citenamefont {Cancio},
  \citenamefont {Chen}, \citenamefont {Krull},\ and\ \citenamefont
  {Burke}}]{Cancio2018}%
  \BibitemOpen
  \bibfield  {author} {\bibinfo {author} {\bibfnamefont {A.}~\bibnamefont
  {Cancio}}, \bibinfo {author} {\bibfnamefont {G.~P.}\ \bibnamefont {Chen}},
  \bibinfo {author} {\bibfnamefont {B.~T.}\ \bibnamefont {Krull}},\ and\
  \bibinfo {author} {\bibfnamefont {K.}~\bibnamefont {Burke}},\ }\bibfield
  {title} {\bibinfo {title} {{Fitting a round peg into a round hole:
  Asymptotically correcting the generalized gradient approximation for
  correlation}},\ }\href {https://doi.org/10.1063/1.5021597} {\bibfield
  {journal} {\bibinfo  {journal} {Journal of Chemical Physics}\ }\textbf
  {\bibinfo {volume} {149}},\ \bibinfo {pages} {084116} (\bibinfo {year}
  {2018})}\BibitemShut {NoStop}%
\bibitem [{\citenamefont {Tao}\ \emph {et~al.}(2003)\citenamefont {Tao},
  \citenamefont {Perdew}, \citenamefont {Staroverov},\ and\ \citenamefont
  {Scuseria}}]{Tao2003}%
  \BibitemOpen
  \bibfield  {author} {\bibinfo {author} {\bibfnamefont {J.}~\bibnamefont
  {Tao}}, \bibinfo {author} {\bibfnamefont {J.~P.}\ \bibnamefont {Perdew}},
  \bibinfo {author} {\bibfnamefont {V.~N.}\ \bibnamefont {Staroverov}},\ and\
  \bibinfo {author} {\bibfnamefont {G.~E.}\ \bibnamefont {Scuseria}},\
  }\bibfield  {title} {\bibinfo {title} {{Climbing the Density Functional
  Ladder: Non-Empirical Meta-Generalized Gradient Approximation Designed for
  Molecules and Solids}},\ }\href
  {https://doi.org/10.1103/PhysRevLett.91.146401} {\bibfield  {journal}
  {\bibinfo  {journal} {Physical Review Letters}\ }\textbf {\bibinfo {volume}
  {91}},\ \bibinfo {pages} {146401} (\bibinfo {year} {2003})}\BibitemShut
  {NoStop}%
\bibitem [{\citenamefont {Sun}\ \emph {et~al.}(2015{\natexlab{a}})\citenamefont
  {Sun}, \citenamefont {Ruzsinszky},\ and\ \citenamefont {Perdew}}]{Sun2015}%
  \BibitemOpen
  \bibfield  {author} {\bibinfo {author} {\bibfnamefont {J.}~\bibnamefont
  {Sun}}, \bibinfo {author} {\bibfnamefont {A.}~\bibnamefont {Ruzsinszky}},\
  and\ \bibinfo {author} {\bibfnamefont {J.~P.}\ \bibnamefont {Perdew}},\
  }\bibfield  {title} {\bibinfo {title} {{Strongly Constrained and
  Appropriately Normed Semilocal Density Functional}},\ }\href
  {https://doi.org/10.1103/PhysRevLett.115.036402} {\bibfield  {journal}
  {\bibinfo  {journal} {Physical Review Letters}\ }\textbf {\bibinfo {volume}
  {115}},\ \bibinfo {pages} {036402} (\bibinfo {year}
  {2015}{\natexlab{a}})}\BibitemShut {NoStop}%
\bibitem [{\citenamefont {Becke}\ and\ \citenamefont
  {Roussel}(1989)}]{Becke1989}%
  \BibitemOpen
  \bibfield  {author} {\bibinfo {author} {\bibfnamefont {A.~D.}\ \bibnamefont
  {Becke}}\ and\ \bibinfo {author} {\bibfnamefont {M.~R.}\ \bibnamefont
  {Roussel}},\ }\bibfield  {title} {\bibinfo {title} {{Exchange holes in
  inhomogeneous systems: A coordinate-space model}},\ }\href
  {https://doi.org/10.1103/PhysRevA.39.3761} {\bibfield  {journal} {\bibinfo
  {journal} {Physical Review A}\ }\textbf {\bibinfo {volume} {39}},\ \bibinfo
  {pages} {3761} (\bibinfo {year} {1989})}\BibitemShut {NoStop}%
\bibitem [{\citenamefont {Tao}\ and\ \citenamefont {Mo}(2016)}]{Tao2016}%
  \BibitemOpen
  \bibfield  {author} {\bibinfo {author} {\bibfnamefont {J.}~\bibnamefont
  {Tao}}\ and\ \bibinfo {author} {\bibfnamefont {Y.}~\bibnamefont {Mo}},\
  }\bibfield  {title} {\bibinfo {title} {{Accurate Semilocal Density Functional
  for Condensed-Matter Physics and Quantum Chemistry}},\ }\href
  {https://doi.org/10.1103/PhysRevLett.117.073001} {\bibfield  {journal}
  {\bibinfo  {journal} {Physical Review Letters}\ }\textbf {\bibinfo {volume}
  {117}},\ \bibinfo {pages} {073001} (\bibinfo {year} {2016})}\BibitemShut
  {NoStop}%
\bibitem [{\citenamefont {Perdew}\ \emph {et~al.}(2007)\citenamefont {Perdew},
  \citenamefont {Ruzsinszky}, \citenamefont {Csonka}, \citenamefont {Vydrov},
  \citenamefont {Scuseria}, \citenamefont {Staroverov},\ and\ \citenamefont
  {Tao}}]{Perdew2007}%
  \BibitemOpen
  \bibfield  {author} {\bibinfo {author} {\bibfnamefont {J.~P.}\ \bibnamefont
  {Perdew}}, \bibinfo {author} {\bibfnamefont {A.}~\bibnamefont {Ruzsinszky}},
  \bibinfo {author} {\bibfnamefont {G.~I.}\ \bibnamefont {Csonka}}, \bibinfo
  {author} {\bibfnamefont {O.~A.}\ \bibnamefont {Vydrov}}, \bibinfo {author}
  {\bibfnamefont {G.~E.}\ \bibnamefont {Scuseria}}, \bibinfo {author}
  {\bibfnamefont {V.~N.}\ \bibnamefont {Staroverov}},\ and\ \bibinfo {author}
  {\bibfnamefont {J.}~\bibnamefont {Tao}},\ }\bibfield  {title} {\bibinfo
  {title} {{Exchange and correlation in open systems of fluctuating electron
  number}},\ }\href {https://doi.org/10.1103/PhysRevA.76.040501} {\bibfield
  {journal} {\bibinfo  {journal} {Physical Review A}\ }\textbf {\bibinfo
  {volume} {76}},\ \bibinfo {pages} {040501} (\bibinfo {year} {2007})},\
  \Eprint {https://arxiv.org/abs/0702283} {arXiv:0702283 [cond-mat]}
  \BibitemShut {NoStop}%
\bibitem [{\citenamefont {Zhao}\ and\ \citenamefont
  {Truhlar}(2006)}]{Zhao2006}%
  \BibitemOpen
  \bibfield  {author} {\bibinfo {author} {\bibfnamefont {Y.}~\bibnamefont
  {Zhao}}\ and\ \bibinfo {author} {\bibfnamefont {D.~G.}\ \bibnamefont
  {Truhlar}},\ }\bibfield  {title} {\bibinfo {title} {{A new local density
  functional for main-group thermochemistry, transition metal bonding,
  thermochemical kinetics, and noncovalent interactions}},\ }\href
  {https://doi.org/10.1063/1.2370993} {\bibfield  {journal} {\bibinfo
  {journal} {Journal of Chemical Physics}\ }\textbf {\bibinfo {volume} {125}},\
  \bibinfo {pages} {194101} (\bibinfo {year} {2006})}\BibitemShut {NoStop}%
\bibitem [{\citenamefont {Furness}\ and\ \citenamefont
  {Sun}(2019)}]{Furness2019}%
  \BibitemOpen
  \bibfield  {author} {\bibinfo {author} {\bibfnamefont {J.~W.}\ \bibnamefont
  {Furness}}\ and\ \bibinfo {author} {\bibfnamefont {J.}~\bibnamefont {Sun}},\
  }\bibfield  {title} {\bibinfo {title} {{Enhancing the efficiency of density
  functionals with an improved iso-orbital indicator}},\ }\href
  {https://doi.org/10.1103/PhysRevB.99.041119} {\bibfield  {journal} {\bibinfo
  {journal} {Physical Review B}\ }\textbf {\bibinfo {volume} {99}},\ \bibinfo
  {pages} {041119} (\bibinfo {year} {2019})}\BibitemShut {NoStop}%
\bibitem [{\citenamefont {Perdew}\ \emph {et~al.}(2004)\citenamefont {Perdew},
  \citenamefont {Tao}, \citenamefont {Staroverov},\ and\ \citenamefont
  {Scuseria}}]{Perdew2004}%
  \BibitemOpen
  \bibfield  {author} {\bibinfo {author} {\bibfnamefont {J.~P.}\ \bibnamefont
  {Perdew}}, \bibinfo {author} {\bibfnamefont {J.}~\bibnamefont {Tao}},
  \bibinfo {author} {\bibfnamefont {V.~N.}\ \bibnamefont {Staroverov}},\ and\
  \bibinfo {author} {\bibfnamefont {G.~E.}\ \bibnamefont {Scuseria}},\
  }\bibfield  {title} {\bibinfo {title} {{Meta-generalized gradient
  approximation: Explanation of a realistic nonempirical density functional}},\
  }\href {https://doi.org/10.1063/1.1665298} {\bibfield  {journal} {\bibinfo
  {journal} {Journal of Chemical Physics}\ }\textbf {\bibinfo {volume} {120}},\
  \bibinfo {pages} {6898} (\bibinfo {year} {2004})}\BibitemShut {NoStop}%
\bibitem [{\citenamefont {Ruzsinszky}\ \emph {et~al.}(2012)\citenamefont
  {Ruzsinszky}, \citenamefont {Sun}, \citenamefont {Xiao},\ and\ \citenamefont
  {Csonka}}]{Ruzsinszky2012}%
  \BibitemOpen
  \bibfield  {author} {\bibinfo {author} {\bibfnamefont {A.}~\bibnamefont
  {Ruzsinszky}}, \bibinfo {author} {\bibfnamefont {J.}~\bibnamefont {Sun}},
  \bibinfo {author} {\bibfnamefont {B.}~\bibnamefont {Xiao}},\ and\ \bibinfo
  {author} {\bibfnamefont {G.~I.}\ \bibnamefont {Csonka}},\ }\bibfield  {title}
  {\bibinfo {title} {{A meta-GGA made free of the order of limits anomaly}},\
  }\href {https://doi.org/10.1021/ct300269u} {\bibfield  {journal} {\bibinfo
  {journal} {Journal of Chemical Theory and Computation}\ }\textbf {\bibinfo
  {volume} {8}},\ \bibinfo {pages} {2078} (\bibinfo {year} {2012})}\BibitemShut
  {NoStop}%
\bibitem [{\citenamefont {Xiao}\ \emph {et~al.}(2013)\citenamefont {Xiao},
  \citenamefont {Sun}, \citenamefont {Ruzsinszky}, \citenamefont {Feng},
  \citenamefont {Haunschild}, \citenamefont {Scuseria},\ and\ \citenamefont
  {Perdew}}]{Xiao2013}%
  \BibitemOpen
  \bibfield  {author} {\bibinfo {author} {\bibfnamefont {B.}~\bibnamefont
  {Xiao}}, \bibinfo {author} {\bibfnamefont {J.}~\bibnamefont {Sun}}, \bibinfo
  {author} {\bibfnamefont {A.}~\bibnamefont {Ruzsinszky}}, \bibinfo {author}
  {\bibfnamefont {J.}~\bibnamefont {Feng}}, \bibinfo {author} {\bibfnamefont
  {R.}~\bibnamefont {Haunschild}}, \bibinfo {author} {\bibfnamefont {G.~E.}\
  \bibnamefont {Scuseria}},\ and\ \bibinfo {author} {\bibfnamefont {J.~P.}\
  \bibnamefont {Perdew}},\ }\bibfield  {title} {\bibinfo {title} {{Testing
  density functionals for structural phase transitions of solids under
  pressure: Si, SiO2, and Zr}},\ }\bibfield  {journal} {\bibinfo  {journal}
  {Physical Review B}\ }\textbf {\bibinfo {volume} {88}},\ \href
  {https://doi.org/10.1103/PhysRevB.88.184103} {10.1103/PhysRevB.88.184103}
  (\bibinfo {year} {2013})\BibitemShut {NoStop}%
\bibitem [{\citenamefont {Kendall}\ \emph {et~al.}(1992)\citenamefont
  {Kendall}, \citenamefont {Dunning},\ and\ \citenamefont
  {Harrison}}]{Kendall1992}%
  \BibitemOpen
  \bibfield  {author} {\bibinfo {author} {\bibfnamefont {R.~A.}\ \bibnamefont
  {Kendall}}, \bibinfo {author} {\bibfnamefont {T.~H.}\ \bibnamefont
  {Dunning}},\ and\ \bibinfo {author} {\bibfnamefont {R.~J.}\ \bibnamefont
  {Harrison}},\ }\bibfield  {title} {\bibinfo {title} {{Electron affinities of
  the first‐row atoms revisited. Systematic basis sets and wave functions}},\
  }\href {https://doi.org/10.1063/1.462569} {\bibfield  {journal} {\bibinfo
  {journal} {The Journal of Chemical Physics}\ }\textbf {\bibinfo {volume}
  {96}},\ \bibinfo {pages} {6796} (\bibinfo {year} {1992})}\BibitemShut
  {NoStop}%
\bibitem [{\citenamefont {Sengupta}\ \emph {et~al.}(2018)\citenamefont
  {Sengupta}, \citenamefont {Bates},\ and\ \citenamefont
  {Ruzsinszky}}]{Sengupta2018}%
  \BibitemOpen
  \bibfield  {author} {\bibinfo {author} {\bibfnamefont {N.}~\bibnamefont
  {Sengupta}}, \bibinfo {author} {\bibfnamefont {J.~E.}\ \bibnamefont
  {Bates}},\ and\ \bibinfo {author} {\bibfnamefont {A.}~\bibnamefont
  {Ruzsinszky}},\ }\bibfield  {title} {\bibinfo {title} {{From semilocal
  density functionals to random phase approximation renormalized perturbation
  theory: A methodological assessment of structural phase transitions}},\
  }\href {https://doi.org/10.1103/PhysRevB.97.235136} {\bibfield  {journal}
  {\bibinfo  {journal} {Physical Review B}\ }\textbf {\bibinfo {volume} {97}},\
  \bibinfo {pages} {235136} (\bibinfo {year} {2018})}\BibitemShut {NoStop}%
\bibitem [{\citenamefont {Kresse}\ and\ \citenamefont
  {Hafner}(1993)}]{Kresse1993}%
  \BibitemOpen
  \bibfield  {author} {\bibinfo {author} {\bibfnamefont {G.}~\bibnamefont
  {Kresse}}\ and\ \bibinfo {author} {\bibfnamefont {J.}~\bibnamefont
  {Hafner}},\ }\bibfield  {title} {\bibinfo {title} {{Ab initio molecular
  dynamics for open-shell transition metals}},\ }\href
  {https://doi.org/10.1103/PhysRevB.48.13115} {\bibfield  {journal} {\bibinfo
  {journal} {Physical Review B}\ }\textbf {\bibinfo {volume} {48}},\ \bibinfo
  {pages} {13115} (\bibinfo {year} {1993})}\BibitemShut {NoStop}%
\bibitem [{\citenamefont {Kresse}\ and\ \citenamefont
  {Furthm{\"{u}}ller}(1996)}]{Kresse1996}%
  \BibitemOpen
  \bibfield  {author} {\bibinfo {author} {\bibfnamefont {G.}~\bibnamefont
  {Kresse}}\ and\ \bibinfo {author} {\bibfnamefont {J.}~\bibnamefont
  {Furthm{\"{u}}ller}},\ }\bibfield  {title} {\bibinfo {title} {{Efficient
  iterative schemes for ab initio total-energy calculations using a plane-wave
  basis set}},\ }\href {https://doi.org/10.1103/PhysRevB.54.11169} {\bibfield
  {journal} {\bibinfo  {journal} {Physical Review B}\ }\textbf {\bibinfo
  {volume} {54}},\ \bibinfo {pages} {11169} (\bibinfo {year}
  {1996})}\BibitemShut {NoStop}%
\bibitem [{\citenamefont {Jain}\ \emph {et~al.}(2013)\citenamefont {Jain},
  \citenamefont {Ong}, \citenamefont {Hautier}, \citenamefont {Chen},
  \citenamefont {Richards}, \citenamefont {Dacek}, \citenamefont {Cholia},
  \citenamefont {Gunter}, \citenamefont {Skinner}, \citenamefont {Ceder},\ and\
  \citenamefont {Persson}}]{Jain2013}%
  \BibitemOpen
  \bibfield  {author} {\bibinfo {author} {\bibfnamefont {A.}~\bibnamefont
  {Jain}}, \bibinfo {author} {\bibfnamefont {S.~P.}\ \bibnamefont {Ong}},
  \bibinfo {author} {\bibfnamefont {G.}~\bibnamefont {Hautier}}, \bibinfo
  {author} {\bibfnamefont {W.}~\bibnamefont {Chen}}, \bibinfo {author}
  {\bibfnamefont {W.~D.}\ \bibnamefont {Richards}}, \bibinfo {author}
  {\bibfnamefont {S.}~\bibnamefont {Dacek}}, \bibinfo {author} {\bibfnamefont
  {S.}~\bibnamefont {Cholia}}, \bibinfo {author} {\bibfnamefont
  {D.}~\bibnamefont {Gunter}}, \bibinfo {author} {\bibfnamefont
  {D.}~\bibnamefont {Skinner}}, \bibinfo {author} {\bibfnamefont
  {G.}~\bibnamefont {Ceder}},\ and\ \bibinfo {author} {\bibfnamefont {K.~A.}\
  \bibnamefont {Persson}},\ }\bibfield  {title} {\bibinfo {title} {{Commentary:
  The materials project: A materials genome approach to accelerating materials
  innovation}},\ }\href {https://doi.org/10.1063/1.4812323} {\bibfield
  {journal} {\bibinfo  {journal} {APL Materials}\ }\textbf {\bibinfo {volume}
  {1}},\ \bibinfo {pages} {011002} (\bibinfo {year} {2013})}\BibitemShut
  {NoStop}%
\bibitem [{\citenamefont {McMahon}\ \emph {et~al.}(1994)\citenamefont
  {McMahon}, \citenamefont {Nelmes}, \citenamefont {Wright},\ and\
  \citenamefont {Allan}}]{McMahon1994}%
  \BibitemOpen
  \bibfield  {author} {\bibinfo {author} {\bibfnamefont {M.~I.}\ \bibnamefont
  {McMahon}}, \bibinfo {author} {\bibfnamefont {R.~J.}\ \bibnamefont {Nelmes}},
  \bibinfo {author} {\bibfnamefont {N.~G.}\ \bibnamefont {Wright}},\ and\
  \bibinfo {author} {\bibfnamefont {D.~R.}\ \bibnamefont {Allan}},\ }\bibfield
  {title} {\bibinfo {title} {{Pressure dependence of the Imma phase of
  silicon}},\ }\href {https://doi.org/10.1103/PhysRevB.50.739} {\bibfield
  {journal} {\bibinfo  {journal} {Physical Review B}\ }\textbf {\bibinfo
  {volume} {50}},\ \bibinfo {pages} {739} (\bibinfo {year} {1994})}\BibitemShut
  {NoStop}%
\bibitem [{\citenamefont {Mujica}\ \emph {et~al.}(2003)\citenamefont {Mujica},
  \citenamefont {Rubio}, \citenamefont {Mu{\~{n}}oz},\ and\ \citenamefont
  {Needs}}]{Mujica2003}%
  \BibitemOpen
  \bibfield  {author} {\bibinfo {author} {\bibfnamefont {A.}~\bibnamefont
  {Mujica}}, \bibinfo {author} {\bibfnamefont {A.}~\bibnamefont {Rubio}},
  \bibinfo {author} {\bibfnamefont {A.}~\bibnamefont {Mu{\~{n}}oz}},\ and\
  \bibinfo {author} {\bibfnamefont {R.~J.}\ \bibnamefont {Needs}},\ }\bibfield
  {title} {\bibinfo {title} {{High-pressure phases of group-IV, III-V, and
  II-VI compounds}},\ }\href {https://doi.org/10.1103/RevModPhys.75.863}
  {\bibfield  {journal} {\bibinfo  {journal} {Reviews of Modern Physics}\
  }\textbf {\bibinfo {volume} {75}},\ \bibinfo {pages} {863} (\bibinfo {year}
  {2003})}\BibitemShut {NoStop}%
\bibitem [{\citenamefont {Hamann}(1996)}]{Hamann1996}%
  \BibitemOpen
  \bibfield  {author} {\bibinfo {author} {\bibfnamefont {D.~R.}\ \bibnamefont
  {Hamann}},\ }\bibfield  {title} {\bibinfo {title} {{Generalized gradient
  theory for silica phase transitions}},\ }\href
  {https://doi.org/10.1103/PhysRevLett.76.660} {\bibfield  {journal} {\bibinfo
  {journal} {Physical Review Letters}\ }\textbf {\bibinfo {volume} {76}},\
  \bibinfo {pages} {660} (\bibinfo {year} {1996})}\BibitemShut {NoStop}%
\bibitem [{\citenamefont {Sun}\ \emph {et~al.}(2013{\natexlab{a}})\citenamefont
  {Sun}, \citenamefont {Haunschild}, \citenamefont {Xiao}, \citenamefont
  {Bulik}, \citenamefont {Scuseria},\ and\ \citenamefont {Perdew}}]{Sun2013}%
  \BibitemOpen
  \bibfield  {author} {\bibinfo {author} {\bibfnamefont {J.}~\bibnamefont
  {Sun}}, \bibinfo {author} {\bibfnamefont {R.}~\bibnamefont {Haunschild}},
  \bibinfo {author} {\bibfnamefont {B.}~\bibnamefont {Xiao}}, \bibinfo {author}
  {\bibfnamefont {I.~W.}\ \bibnamefont {Bulik}}, \bibinfo {author}
  {\bibfnamefont {G.~E.}\ \bibnamefont {Scuseria}},\ and\ \bibinfo {author}
  {\bibfnamefont {J.~P.}\ \bibnamefont {Perdew}},\ }\bibfield  {title}
  {\bibinfo {title} {{Semilocal and hybrid meta-generalized gradient
  approximations based on the understanding of the kinetic-energy-density
  dependence Semilocal and hybrid meta-generalized gradient approximations
  based on the understanding of the kinetic-energy-density depend}},\ }\href
  {https://doi.org/10.1063/1.4789414} {\bibfield  {journal} {\bibinfo
  {journal} {Journal of Chemical Physics}\ }\textbf {\bibinfo {volume} {138}},\
  \bibinfo {pages} {044113} (\bibinfo {year} {2013}{\natexlab{a}})}\BibitemShut
  {NoStop}%
\bibitem [{\citenamefont {Sun}\ \emph {et~al.}(2013{\natexlab{b}})\citenamefont
  {Sun}, \citenamefont {Xiao}, \citenamefont {Fang}, \citenamefont
  {Haunschild}, \citenamefont {Hao}, \citenamefont {Ruzsinszky}, \citenamefont
  {Csonka}, \citenamefont {Scuseria},\ and\ \citenamefont {Perdew}}]{Sun2013a}%
  \BibitemOpen
  \bibfield  {author} {\bibinfo {author} {\bibfnamefont {J.}~\bibnamefont
  {Sun}}, \bibinfo {author} {\bibfnamefont {B.}~\bibnamefont {Xiao}}, \bibinfo
  {author} {\bibfnamefont {Y.}~\bibnamefont {Fang}}, \bibinfo {author}
  {\bibfnamefont {R.}~\bibnamefont {Haunschild}}, \bibinfo {author}
  {\bibfnamefont {P.}~\bibnamefont {Hao}}, \bibinfo {author} {\bibfnamefont
  {A.}~\bibnamefont {Ruzsinszky}}, \bibinfo {author} {\bibfnamefont {G.~I.}\
  \bibnamefont {Csonka}}, \bibinfo {author} {\bibfnamefont {G.~E.}\
  \bibnamefont {Scuseria}},\ and\ \bibinfo {author} {\bibfnamefont {J.~P.}\
  \bibnamefont {Perdew}},\ }\bibfield  {title} {\bibinfo {title} {{Density
  functionals that recognize covalent, metallic, and weak bonds}},\ }\href
  {https://doi.org/10.1103/PhysRevLett.111.106401} {\bibfield  {journal}
  {\bibinfo  {journal} {Physical Review Letters}\ }\textbf {\bibinfo {volume}
  {111}},\ \bibinfo {pages} {106401} (\bibinfo {year}
  {2013}{\natexlab{b}})}\BibitemShut {NoStop}%
\bibitem [{\citenamefont {Sun}\ \emph {et~al.}(2015{\natexlab{b}})\citenamefont
  {Sun}, \citenamefont {Perdew},\ and\ \citenamefont {Ruzsinszky}}]{Sun2015a}%
  \BibitemOpen
  \bibfield  {author} {\bibinfo {author} {\bibfnamefont {J.}~\bibnamefont
  {Sun}}, \bibinfo {author} {\bibfnamefont {J.~P.}\ \bibnamefont {Perdew}},\
  and\ \bibinfo {author} {\bibfnamefont {A.}~\bibnamefont {Ruzsinszky}},\
  }\bibfield  {title} {\bibinfo {title} {{Semilocal density functional obeying
  a strongly tightened bound for exchange}},\ }\href
  {https://doi.org/10.1073/pnas.1423145112} {\bibfield  {journal} {\bibinfo
  {journal} {Proceedings of the National Academy of Sciences}\ }\textbf
  {\bibinfo {volume} {112}},\ \bibinfo {pages} {685} (\bibinfo {year}
  {2015}{\natexlab{b}})}\BibitemShut {NoStop}%
\bibitem [{\citenamefont {Hao}\ \emph {et~al.}(2012)\citenamefont {Hao},
  \citenamefont {Fang}, \citenamefont {Sun}, \citenamefont {Csonka},
  \citenamefont {Philipsen},\ and\ \citenamefont {Perdew}}]{Hao2012}%
  \BibitemOpen
  \bibfield  {author} {\bibinfo {author} {\bibfnamefont {P.}~\bibnamefont
  {Hao}}, \bibinfo {author} {\bibfnamefont {Y.}~\bibnamefont {Fang}}, \bibinfo
  {author} {\bibfnamefont {J.}~\bibnamefont {Sun}}, \bibinfo {author}
  {\bibfnamefont {G.~I.}\ \bibnamefont {Csonka}}, \bibinfo {author}
  {\bibfnamefont {P.~H.~T.}\ \bibnamefont {Philipsen}},\ and\ \bibinfo {author}
  {\bibfnamefont {J.~P.}\ \bibnamefont {Perdew}},\ }\bibfield  {title}
  {\bibinfo {title} {{Lattice constants from semilocal density functionals with
  zero-point phonon correction}},\ }\href
  {https://doi.org/10.1103/PhysRevB.85.014111} {\bibfield  {journal} {\bibinfo
  {journal} {Physical Review B}\ }\textbf {\bibinfo {volume} {85}},\ \bibinfo
  {pages} {014111} (\bibinfo {year} {2012})}\BibitemShut {NoStop}%
\end{thebibliography}%

\end{document}